\documentclass[aps,10pt,prl,reprint,twocolumn,floatfix,showpacs,superscriptaddress]{revtex4-1}  

\usepackage{amsfonts,amsmath,amssymb}
\usepackage[margin=20mm,inner=20mm,marginpar=10mm]{geometry}
\usepackage{graphicx, epstopdf}
\usepackage{microtype}
\usepackage[figuresright]{rotating}
\usepackage{setspace}
\usepackage{textcomp}
\usepackage{times}
\usepackage[normalem]{ulem}
\usepackage[abs]{overpic}
\usepackage{color}
\usepackage[T1]{fontenc}
\usepackage{physics}

\newlength{\figwidth}
\begin{document}

\title{A fiber-based beam profiler for high-power laser beams in confined spaces and ultra-high vacuum}

\author{Christian Brand}
\email{brandc6@univie.ac.at}
\affiliation{University of Vienna, Faculty of Physics, Boltzmanngasse 5, A-1090 Vienna, Austria}
\affiliation{German Aerospace Center (DLR), Institute of Quantum Technologies, S\"oflinger Stra\ss e 100, 89077 Ulm, Germany}

\author{Ksenija Simonovi\'c}
\affiliation{University of Vienna, Faculty of Physics, Boltzmanngasse 5, A-1090 Vienna, Austria}

\author{Filip Kia{\l}ka}
\affiliation{University of Vienna, Faculty of Physics, Boltzmanngasse 5, A-1090 Vienna, Austria}

\author{Stephan Troyer}
\affiliation{University of Vienna, Faculty of Physics, Boltzmanngasse 5, A-1090 Vienna, Austria}

\author{Philipp Geyer}
\affiliation{University of Vienna, Faculty of Physics, Boltzmanngasse 5, A-1090 Vienna, Austria}

\author{Markus Arndt}
\affiliation{University of Vienna, Faculty of Physics, Boltzmanngasse 5, A-1090 Vienna, Austria}
	
\date{\today}

\begin{abstract}
Laser beam profilometry is an important scientific task with well-established solutions for beams propagating in air. It has, however, remained an open challenge to measure beam profiles of high-power lasers in ultra-high vacuum and in tightly confined spaces. Here we present a novel scheme that uses a single multi-mode fiber to scatter light and guide it to a detector. The method competes well with commercial systems in position resolution, can reach through apertures smaller than $500\times 500$~$\mu$m$^2$ and is compatible with ultra-high vacuum conditions. The scheme is simple, compact, reliable and can withstand laser intensities beyond 2~MW/cm$^2$.
\end{abstract}

\maketitle

\section{Introduction}

Since the advent of the laser~\cite{Maiman_Nature187_493} in 1960, a variety of techniques have been developed to precisely characterize transverse beam profiles~\cite{Yariv_OpticalElectronics,Eichler_Lasers2018}. These may be Gaussian or flat-top, round or rectangular, symmetric or asymmetric. At low intensities and for visible to near-infrared wavelengths, two-dimensional high-resolution images have been recorded using photographic techniques~\cite{Milam_ApplOpt20_169} that can even be self-calibrating~\cite{Winer_ApplOpt5_1437}. Laser profilometry by photothermal deflection~\cite{Rose_ApplOpt25_1738} and thermography~\cite{Baba_RevSciInstrum57_2739} are viable options, but hardly generalizable to arbitrary beam shapes. The use of microelectromechanical systems (MEMS)~\cite{Sumriddetchkajorn_ApplOpt41_3506} and surface-plasmon-polaritons~\cite{Dietbacher_OptExpress29_1408} has been successfully demonstrated in profilometry, even down to sub-$\mu$m resolution, but with some constraints in laser intensity.

Nowadays two-dimensional images are obtained in high quality using silicon-based photodiode arrays~\cite{Liu_OptLett7_196}, CCD and CMOS cameras or even webcams and smartphones~\cite{Hossain_OptLett40_5156}, which can be sensitive megapixel detectors with a pixel size down to 5~$\mu$m. However, these elements start to be irreversibly damaged at laser intensities above $10$~kW/cm$^2$~\cite{Schwarz_OptEng56_034108}. Furthermore, near-infrared radiation beyond 1.1~$\mu$m remains undetected because it falls into the energy gap of silicon. Alternative materials such as InGaAs are available, but chips with high resolution are still challenging to produce and have similar damage thresholds.
 
To characterize high-intensity laser beams usually rotating knife-edges~\cite{Suzaki_ApplOpt14_2809,Veshapidze_ApplOpt45_8197}, scanning slits~\cite{Soto_ApplOpt36_7450} and scanning pinholes~\cite{Sheikh_IEEEPhotoicsTechnolLett21_666} are used. The spatial resolution is then provided by a mechanical element, which can sustain high energies. Two-dimensional images are reconstructed by scanning the obstacle and the final resolution is defined by the sharpness of the edge, slit, or hole. All of the above-mentioned processes and techniques work very well on an optical table in air and have found many experimental and commercial realizations.
 
Here we explore a complementary domain of applications, motivated by molecular matter-wave interferometry~\cite{Arndt_NPhys10_271,Nairz_PRL87_160401,Gerlich_NPhys3_711,Haslinger_NaturePhys9_144,Brand_NatNanotechnol10_845}, where complex molecules are diffracted at nanomechanical gratings and standing light waves. Understanding the interaction of matter-waves with light gratings requires the precise knowledge of the laser intensity in immediate proximity (several tens of $\mu$m) to a highly reflective mirror. Moreover, it is a common requirement to determine the beam shape for $30$~W of continuous laser power focused  in high or ultra-high vacuum, typically below $10^{-8}$~mbar. None of the commercial or previously demonstrated detectors could match all these conditions.

Our solution makes use of a cut multi-mode quartz fiber, which scatters light at its edge and guides it to a photodetector.
The described solution is simple, affordable, and fulfills the above-mentioned requirements.
Furthermore, the fiber geometry allows for 2D scanning and for aligning of multiple intersecting lasers with micrometer accuracy.
Being a mechanical obstacle, the same fiber can also be used to align an atomic or molecular beam and its overlap with a collinear or crossing laser beam.
These are key advantages over existing approaches.
 
\section{Concept and Theory}

\begin{figure}[tb]
\includegraphics[width=\linewidth]{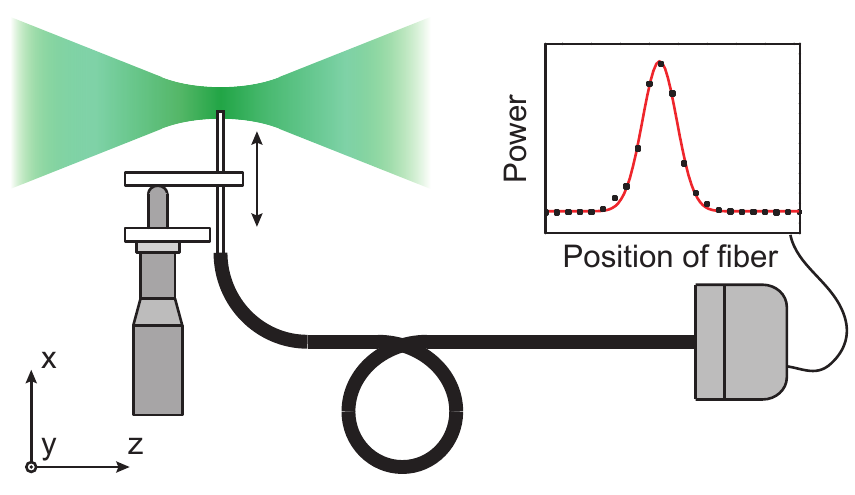}
\caption{A fiber tip is moved with $\mu$m-precision into a focused laser beam and the amount of light scattered into the fiber is registered with a power meter.}
\label{fig:Setup}
\end{figure}

\begin{figure}[htb]
\includegraphics[width=\linewidth]{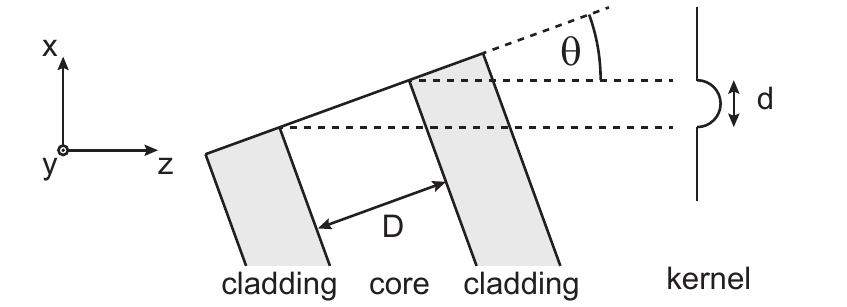}
\caption{For a flat-top fiber tip held at an angle $\theta$ the convolution kernel can be approximated as the projection of the core size $D$ along the beam axis $z$.}
\label{fig:Coupling}
\end{figure}

The concept of the beam profiler is shown in Fig.~\ref{fig:Setup}.
The tip of a stripped multi-mode fiber is moved through a laser beam and scatters light at the core-air interface depending on the local laser intensity.
Part of this light is then guided by the fiber to a suitable detector where it is registered.
High-resolution beam profiling requires that the region from which light is collected is small along the measurement direction ($x$).
One way to achieve this is to use a fiber with a very small core.
Another option is to take a fiber with larger core diameter, produce a flat cut surface and position the tip so that the projection of the cut along the laser beam is small.
In either case we expect the signal to be approximately given by a convolution of the laser intensity with an elliptical kernel
\begin{equation} \label{eq:kernel}
K=	\Re \sqrt{\left(\frac{d}{2}\right)^2 - x^2},
\end{equation}
where $d$ is bounded by the core diameter $D$.
For a flat-tip cut making an angle $\theta$ with the beam axis, we can estimate $d \approx D\sin \theta$, as shown in Fig.~\ref{fig:Coupling}.

To illustrate the potential of this method, we consider a fiber with a core size of 50~$\mu$m and an angle $\theta = 10^\circ$.
The resulting effective detector width of $d \approx$ 8.6~$\mu$m leads to overestimating the width of a laser beam with a 1/e$^2$ intensity radius of 10~$\mu$m by only 1~$\mu$m
For wider beams, the relative error is even smaller, suggesting that high-resolution beam profiling is possible with this method even with moderate alignment and core-size requirements.
Moreover, this simple model assumes that scattering is uniform over the whole surface of the core.
It thus represents an upper limit on the signal broadening, which can be smaller if scattering occurs only at a part of the core surface.
\section{Experimental details}

\begin{figure}[htb]
\includegraphics[width=\linewidth]{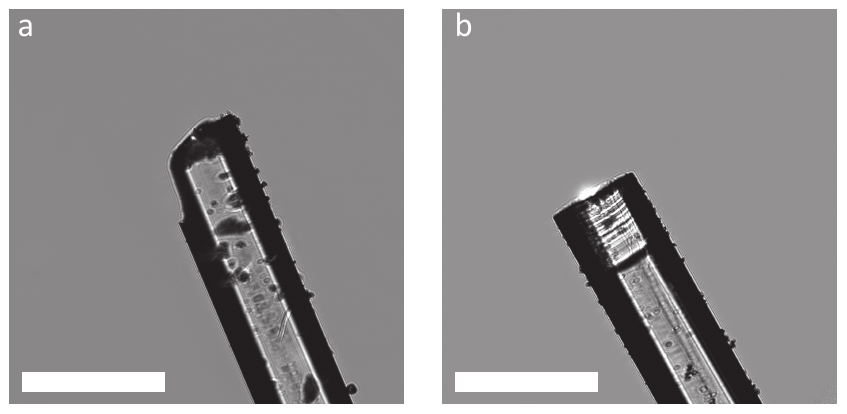}
\caption{
	Microscope images of typical fiber tips used.
	The core size amounts to 10~$\mu$m (a) and 50~$\mu$m (b).
	While the 10~$\mu$m core fiber was cut with pincers, the 50~$\mu$m core fiber was prepared using a scribe and abrasive paper.
	The bright areas show that light is scattered out of the fiber at right angles.
	The bar corresponds to 200~$\mu$m.
}
\label{fig:Fibres}
\end{figure}

To test the model we used multi-mode fibers with a core diameter of 10~$\mu$m (Thorlabs M65L01) and 50~$\mu$m (Thorlabs M16L01).
The 10~$\mu$m core fiber is designed for wavelengths of 400-550~nm and 700-1000~nm, while the other one supports light in the wavelength range of 400-2400~nm.
Both fibers were stripped at their tip, leaving only the polymer cladding and the quartz core.
The 10~$\mu$m core fiber was cut coarsely with a pair of pincers, while the 50~$\mu$m core fiber was cut with either a fiber scribe or a cleaver to produce a flat tip.
Since no light could be coupled at right angles into the pristine cleaved fibers, we roughened their tips with a 30~$\mu$m grain abrasive paper.
This created a sufficient number of scattering centers while maintaining a flat tip.
To assess the quality of the tips and the size of the coupling region, we sent light at several wavelengths (400 to 650~nm) backwards through the fibers and observed the tips under a microscope (see Fig.~\ref{fig:Fibres}).

The experimental setup consists of a 532~nm laser beam (Coherent Verdi V18) expanded to a width of $w_{x,y}=7$~mm, which is vertically focused along the $x$-direction using a cylindrical lens with a focal length of $f =250$~mm.
To generate reference data for the fiber-based beam profiler, we limited the power to 7~mW and positioned a moving edge beam profiler (Coherent BeamMaster) at the focus of the beam.
Then, the focusing lens was moved in steps of 0.5~mm along the beam axis ($z$) using a linear stage and the width $w_x$ of the beam was measured at each lens position.
Analogous measurements were performed with the fibers at up to 14~W of laser power.
In these measurements the fibers were pointing either along the $x$-axis or the $y$-axis and were moved into the laser using a linear stage with a step size of 10~$\mu$m, while the lens was moved again in steps of 0.5~mm.
The light that coupled into the fiber was registered using a power meter (Coherent 818-UV) and the resulting trace was fitted with a Gaussian curve.

The high-intensity tests were done in air and in vacuum ($p < 1\times 10^{-7}$~mbar) by focusing the laser with a spherical lens ($f=250$~mm).
In air we monitored the intensity of coupled light over time.
In vacuum we measured the intensity profile of the focused beam at different laser powers to check for potential damages to the fiber tip.

\section{Results and Discussions}

\begin{figure}[t]
\includegraphics[width=\linewidth]{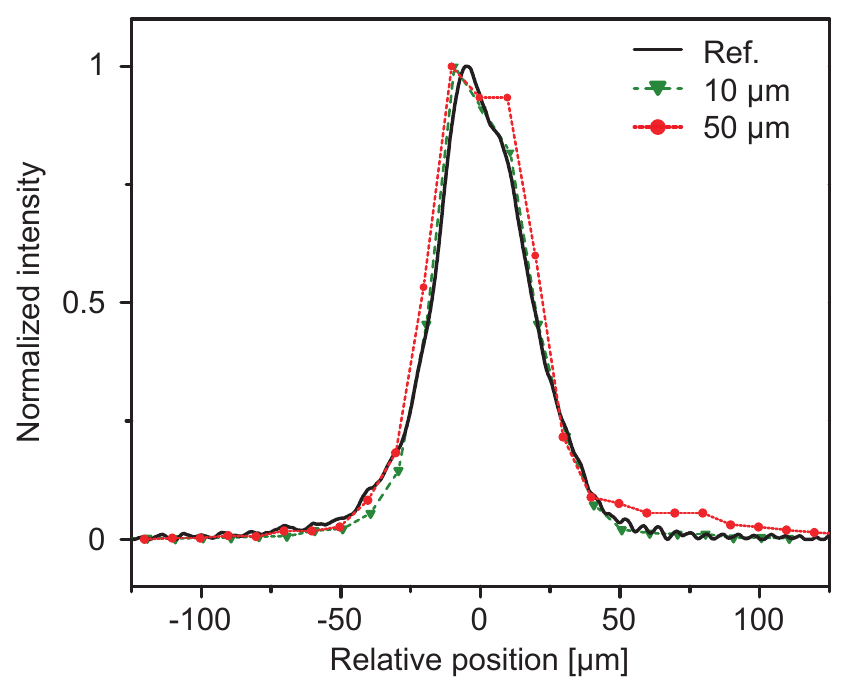}
\caption{Transverse beam profile of a 532~nm laser as obtained with the  moving edge beam profiler (reference) and both fibers.}
\label{fig:Gaussian}
\end{figure}

We start with measuring the width of the laser focused with the cylindrical lens in air.
Fig.~\ref{fig:Gaussian} shows typical beam profiles obtained with both fibers and the moving edge beam profiler.
When the tips cross the beam, light is abundantly scattered into the fiber which allows us to reliably measure the laser beam width.
As desired, the detected light falls back to the background level as soon as the tip exits the beam.
This is the case even if the fiber cladding is still in the beam, corroborating the idea that  light is coupled into the fiber only at the tip.
The respective widths and standard deviations as extracted from a Gaussian fit of the data amount to $32.6\pm1.1~\mu$m for the 10~$\mu$m core fiber and $34.6\pm 1.2~\mu$m for the 50~$\mu$m core fiber.
This is in good agreement with the value obtained with the beam profiler ($33.9\pm 0.1~\mu$m), especially given that the fibers are scanned with a step size of 10~$\mu$m.

\begin{figure}[tb]
\includegraphics[width=\linewidth]{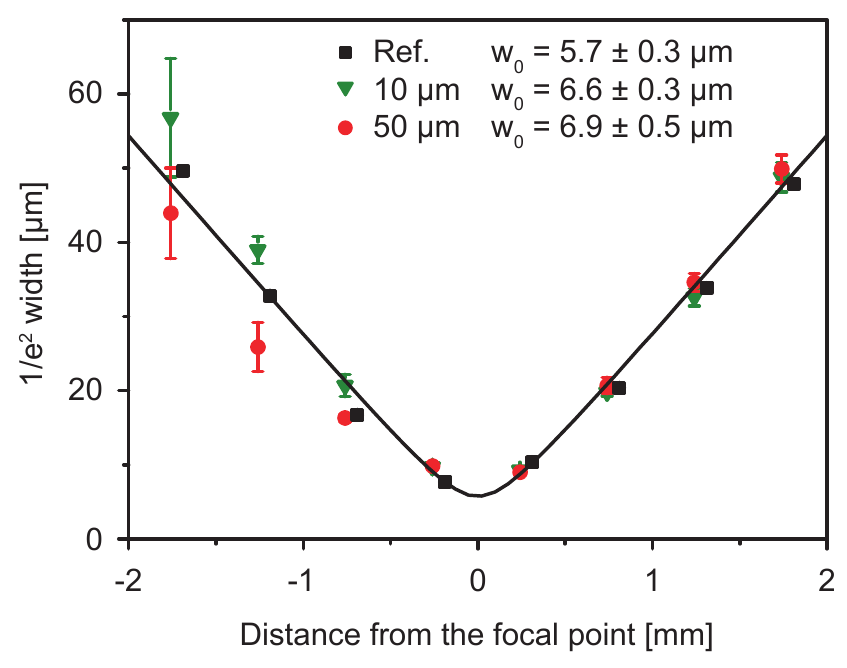}
\caption{
	Comparison of the beam widths extracted from the measurements using fibers with a core size of 10~$\mu$m (green triangles) and 50~$\mu$m (red circles), as well as the moving edge beam profiler (black squares).
	In these measurements the fiber tip is in the $yz$-plane and moved along the $x$-axis.
	All measurements are aligned with respect to the focal point $z_0$. 
	The continuous curve is a hyperbolic fit to the reference data. 
	For distances below -1.5~mm the beam profile deviated from a Gaussian profile, leading to larger error bars in this region.
}
\label{fig:Focus}
\end{figure}

These measurements were repeated at different positions along the beam path and the extracted 1/e$^2$ widths were plotted against the respective distance from the focal point, as shown in Fig.~\ref{fig:Focus}. From the hyperbolic fit of the transverse beam width $w_x$ as a function of longitudinal position $z$ we extracted the focal position $z_0$ and the waist $w_0$ of the beam
\begin{equation}
w_x=w_0\times\sqrt{1+\left(\frac{z-z_0}{z_R}\right)^2}
\label{eqn:width}
\end{equation}
where $z_R$ is the Rayleigh length. 

The reference value of $w_0= 5.7 \pm 0.3~\mu$m as obtained with the moving edge beam profiler is well reproduced both by the 10~$\mu$m ($w_0= 6.6 \pm 0.3~\mu$m) and the 50~$\mu$m core fiber ($w_0= 6.9\pm 0.5~\mu$m).
The overestimation is consistent with the signal being a convolution of the intensity with the kernel~[Eq.~\eqref{eq:kernel}] for $d = 6.4$ ($10$~$\mu$m fiber) and $d = 7.4~\mu$m (50~$\mu$m fiber).
For the 10~$\mu$m fiber the size of the kernel is bounded by the diameter of the core, as expected.
For the 50~$\mu$m fiber the size of the kernel is consistent with the fiber cut making an angle of $8.5^\circ$ with the laser beam axis.
All data were collected in a single run, indicating that this is sufficient to grant high resolution beam profiling. 
In this study the alignment of the fiber with the beam has not been optimized. 
However, in doing so the projection of the core onto the beam axis be can reduce considerably, yielding a kernel which is ultimately only bounded by the roughness of the core surface.

These results illustrate that we can measure the beam waist with a resolution of about 1~$\mu$m, which is substantially smaller than both the fiber core diameters and the scanning step size.
They further indicate that increasing the core diameter from 10 to 50~$\mu$m, and thus enhancing the amount of registered light, does not significantly compromise the precision of the profile measurement.

\begin{figure}[htb]
\includegraphics[width=\linewidth]{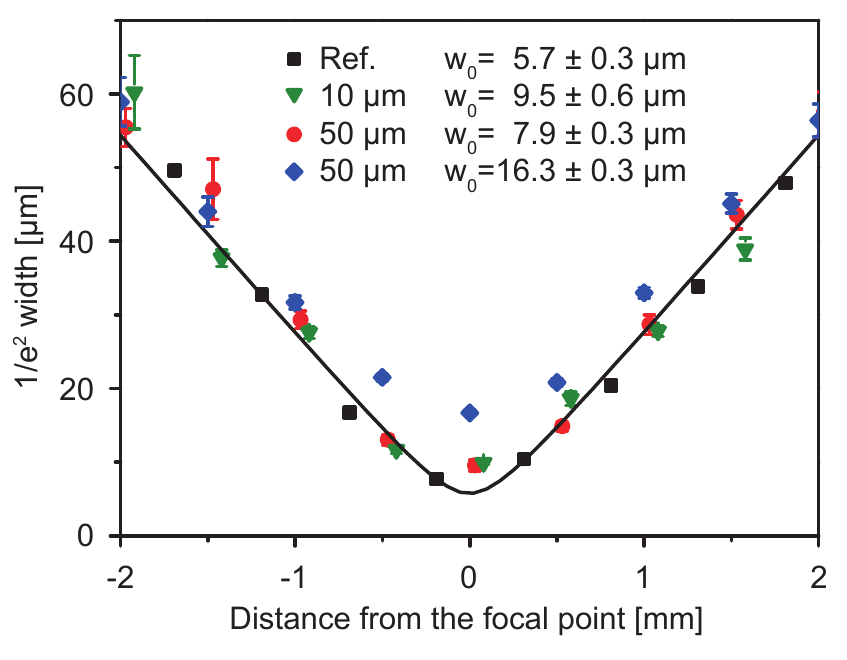}
\caption{
	Comparison of the beam widths extracted from the measurements using fibers with a core size of 10~$\mu$m (green triangles), 50~$\mu$m (red dots and blue rhombuses), and the moving edge beam profiler (black squares).
	In these measurements the fiber tip is in the $xz$-plane and moved along the $x$-axis.
	The continuous curve is a hyperbolic fit to the reference data.
	All measurements are aligned with respect to the focal point $z_0$.
	}
\label{fig:Vertical}
\end{figure}

To investigate the resolution obtained when scanning the fiber transversally, as required for 2D profiling, we repeated the measurements with the fiber pointing along the $y$-direction.
The results, illustrated in Fig.~\ref{fig:Vertical}, show that the 10~$\mu$m fiber gives good agreement with the reference measurements also when scanned transversally.
When using fibers with a core size of 50~$\mu$m, the resolution varies from tip to tip despite the same cutting procedure.
However, the broadening is always below the value resulting from a convolution of the intensity with~Eq.~\eqref{eq:kernel} for $d = 50~\mu$m.

To ensure the fiber is compatible with high power densities, we focused a laser beam at 532~nm with a spherical lens ($f = 250$~mm) from the side onto the 10~$\mu$m fiber tip in air.
The out-coupled power was monitored over time to check for any thermal or optomechanical damages.
It stayed stable for longer than 180 minutes, the maximum time window tested in this series, even at intensities up to 2~MW/cm$^2$.

\begin{figure}[htb]
\includegraphics[width=\linewidth]{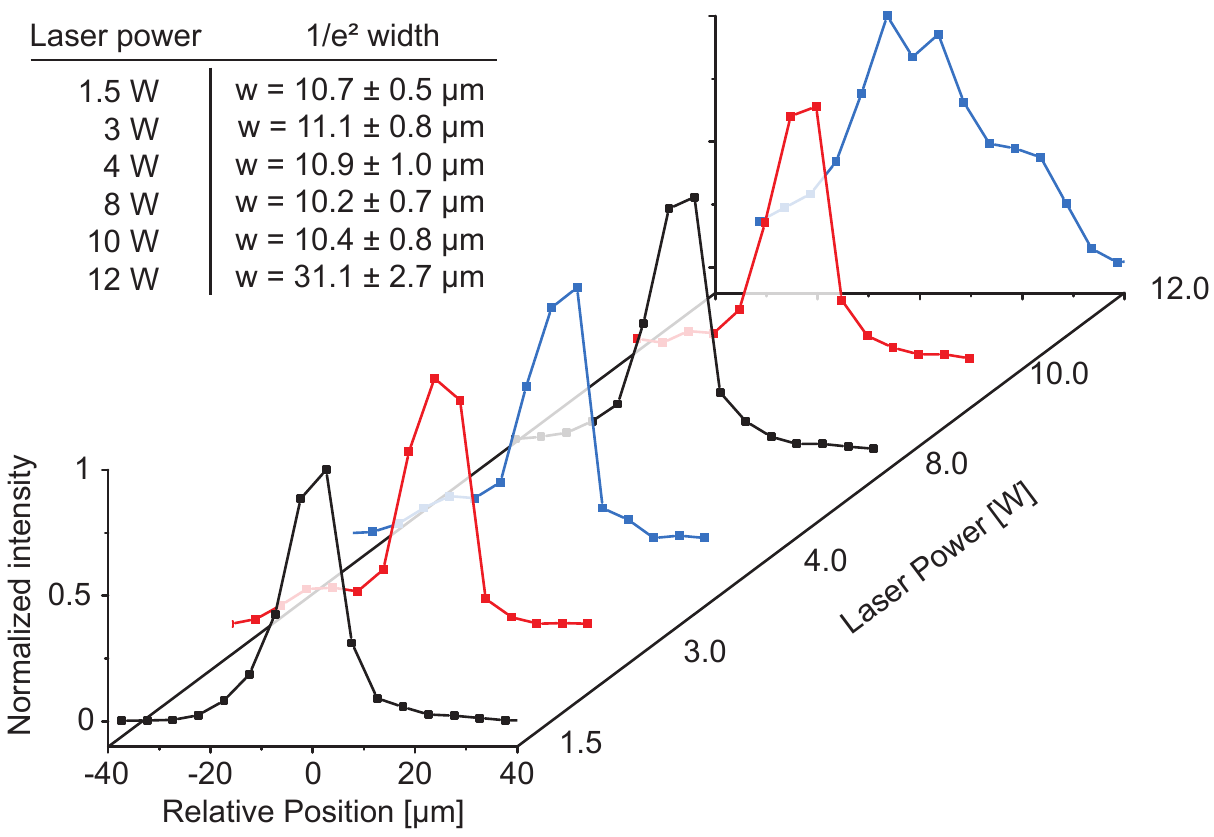}
\caption{
Beam widths for different laser intensities measured with a 50~$\mu$m core fiber in high vacuum.
The width can be measured reliably while increasing the laser power from 1.5 to 10~W. 
The sudden increase in the beam width at 12~W is most likely caused by degradation of the fiber tip.}
\label{fig:vacuum}
\end{figure}

In vacuum we focused the laser onto the tip of a 50~$\mu$m core fiber and increased the power in steps from 1.5 to 12~W.
The waist of the beam was measured at each power after exposing the fiber tip for 5 minutes to the focused laser.
We could reliably measure the width for powers up to 10~W focused to a waist of 10~$\mu$m, corresponding to a peak intensity of more than 4~MW/cm$^2$.
At a power of 12 W, the measured radius suddenly increased to $31\pm 3~\mu$m, as shown in Fig.~\ref{fig:vacuum}, which we assign to thermal degradation of the tip. 
During the vacuum test, the polymer cladding was measurably outgassing at intensities above 200~kW/cm$^2$, limiting the high-vacuum compatibility.
However, this can be mitigated by replacing the polymer-cladded fibers with glass-cladded ones.
We also performed long-term measurements where we could reliably measure the width of a focused laser in high vacuum over a period of three months.

\section{Conclusion}

In conclusion, fiber sampling is shown to be an easy and minimally invasive method to determine laser beam parameters. 
Because the fiber itself has a diameter below 200~$\mu$m and can be scanned on a lever arm, it can be inserted in constricted and otherwise inaccessible places, such as for instance in 100~$\mu$m distance to a mirror or inside an open access microcavity~\cite{Wachter_LightSciAppl8_1,Garcia_OptExpress26_22249} in high vacuum. 
Actuators with nanometer resolution are readily available and allow for automated profile measurements, also in 2D or 3D. 
The method can be applied to laser beams of arbitrary shape and any wavelength that is supported by the fiber.

The technique might potentially be extended to super-resolution as well, with the use of extruded fiber tips such as those used in near-field scanning optical microscopy.
It appears realistic that a resolution of better than $\lambda/10$ can be achieved this way.
Further, it could be extended to pulsed laser beams, again of high intensity.

\section{Acknowledgements} This project has received funding from the Austrian Science Funds (FWF) (P-30176). We acknowledge help from Maxime Debiossac and Jakob Rieser in preparing the fiber cuts.


%

\end{document}